\documentclass[aps,prb,twocolumn,notitlepage,superscriptaddress,showpacs,10pt,longbibliography,floatfix]{revtex4-2} 

\usepackage{amsmath}    
\usepackage{amsfonts}
\usepackage{bm}
\usepackage{amssymb}
\usepackage{graphicx}   
\usepackage[usenames,dvipsnames]{xcolor}      
\usepackage{comment}
\usepackage{siunitx}
\usepackage{enumitem}

\usepackage{soul}
\usepackage[normalem]{ulem}

\mathchardef\mhyphen="2D

\renewcommand{\vec}[1]{\mathbf{#1}}

\begin{document}

\title{Local invariants identify topology in metals and gapless systems}

\author{Alexander Cerjan}
\email[]{awcerja@sandia.gov}
\affiliation{Center for Integrated Nanotechnologies, Sandia National Laboratories, Albuquerque, New Mexico 87185, USA}
\author{Terry A.\ Loring}
\email[]{loring@math.unm.edu}
\affiliation{Department of Mathematics and Statistics, University of New Mexico, Albuquerque, New Mexico 87131, USA}

\date{\today}

\begin{abstract}
    Although topological band theory has been used to discover and classify a wide array of novel topological phases in insulating and semi-metal systems, it is not well-suited to identifying topological phenomena in metallic or gapless systems. Here, we develop a theory of topological metals based on the system's spectral localizer and associated Clifford pseudospectrum, which can both determine whether a system exhibits boundary-localized states despite the presence of degenerate bulk bands and provide a measure of these states' topological protection even in the absence of a bulk band gap. We demonstrate the generality of this method across symmetry classes in two lattice systems, a Chern metal and a higher-order topological metal, and prove the topology of these systems is robust to relatively strong perturbations. The ability to define invariants for metallic and gapless systems allows for the possibility of finding topological phenomena in a broad range of natural, photonic, and other artificial materials that could not be previously explored. 
\end{abstract}

\maketitle

\section{Introduction}

Topological band theory has enabled enormous progress in the discovery and classification of novel states of matter. The preponderance of these developments have been predicted and realized in insulators \cite{su_soliton_1980,klitzing_new_1980,halperin_quantized_1982,thouless_quantized_1982,haldane_model_1988,buttiker_absence_1988,zak_berrys_1989,king1993theory,fang_anomalous_2003,kane_z_2_2005,bernevig_quantum_2006,fu_topological_2007,konig_quantum_2007,qi_topological_2008,haldane_possible_2008,hsieh_topological_2008,wang_observation_2009,hsieh_observation_2009,hsieh_tunable_2009,hafezi_robust_2011,umucalilar_artificial_2011,ringel_strong_2012,kitagawa_observation_2012,fang_realizing_2012,hafezi_imaging_2013,rechtsman_photonic_2013,khanikaev_photonic_2013,benalcazar2017quad,benalcazar2017quadPRB,song_densuremath-2-dimensional_2017,serra2018observation,peterson2018,noh2018,oded2018} and semi-metals \cite{wan_topological_2011,young_dirac_2012,wang_dirac_2012,wang_three-dimensional_2013,narayan_topological_2014,xu_observation_2015,lu_experimental_2015,xu_discovery_2015,chen_photonic_2016,lu_symmetry-protected_2016,lv_observation_2017,ma_direct_2017,kawakami_symmetry-guaranteed_2017,goi_observation_2018,ying_symmetry-protected_2018,schulz2021invariants}, where these systems' topological features are easily identified due to their isolation in energy and wavevector $(E,\vec{k})$-space. However, in metals and other materials lacking a bulk band gap, any states of topological origin are degenerate with bulk states, generally resulting in hybridization between the two sets of states. This hybridization makes it difficult to say whether a particular set of states remains localized to the system's boundaries, or retains any of the other topological properties that they would possess in an insulating system. Moreover, even if boundary-localized states could be identified, the absence of a bulk band gap means that traditional topological band theories would be unable to predict whether these states would be robust to perturbations, or quantify the strength of that protection. Although detailed studies in particular metallic and gapless systems have demonstrated the existence of some topological behaviors \cite{bergman_bulk_2010,bergman_axion_2011,barkeshli_topological_2011,karch_surface_2011,junck_transport_2013,meyer_disordered_2013,benalcazar2020_HOT_BIC_theory,cerjan2020_HOT_BIC_exp,jung_exact_2021}, a general theory for predicting topological phenomena in any metallic or gapless system has remained elusive.

Theories of topological materials predicated upon diagonalizing a system's Hamiltonian to determine its topology possess an inherent challenge when considering metals or other gapless materials. In general, the Hamiltonians, $H$, of topologically non-trivial systems do not commute with position operators, $X$, i.e., $[H,X] \ne 0$. Thus, it is impossible to find eigenstates of both operators simultaneously. In insulators and semi-metals, where any energy eigenstates of topological origin can be spectrally isolated, position expectation values provide a measure of the state's location and localization. However, in metals or gapless systems, any potential topological energy eigenstate is a member of a large degenerate subspace consisting primarily of bulk states, which renders position expectation values meaningless without some other discriminant between possible choices of basis within this subspace. This argument suggests that a theory of topological metals should be pursued using real-space definitions of topology that do not require diagonalizing the Hamiltonian \cite{kitaev2006anyons,bianco_local_chern_2011,Fulga_et_al_scattering_top_ins,loring_Hastings2011disordered,mitchell_amorphous_2018,Varjas_et_al_2020_KMP_method,prodan2011disordered}. Such real-space topological theories have recently been used to identify distinct phases in aperiodic systems, such as quasicrystals \cite{tran_topological_2015,bandres_topological_2016,fulga_aperiodic_2016}, amorphous structures \cite{agarwala_topological_2017,mansha_robust_2017,xiao_photonic_2017,mitchell_amorphous_2018,poyhonen_amorphous_2018,bourne_non-commutative_2018,yang_topological_2019,agarwala_higher-order_2020}, and fractal lattices \cite{yang_photonic_2020}. 

Here, we develop a general theory of topological metals and other gapless materials defined using local invariants derived from the system's ``spectral localizer'' \cite{loringPseuspectra}. This theory has three inherent features that allow it to directly solve the difficulties facing any topological theory of gapless systems: First, as the spectral localizer treats the system's Hamiltonian on equal footing with its position operators, it is able to simultaneously identify the approximate energy and position of the system's states. Second, in the absence of a state, the spectral localizer returns a measure of the strength of the perturbation required to move a state to that position and energy --- in particular, one can calculate the Clifford pseudospectrum (a set defined by the spectral localizer) in the immediate vicinity of a boundary-localized state to determine the strength of its protection against disorder. 
Third, the spectral localizer is mathematically proven to be connected to the system's $K$-theory and thus can be used to define local invariants that classify the system's topological phase at a given energy and position \cite{loringPseuspectra,LoringSchuBa_even, LoringSchuBa_odd}. To demonstrate the generality of this method, we explicitly determine the topological character and quantify the strength of its protection in two disparate models, a Chern metal and a higher-order topological metal. 
As part of this study, we also provide a definition of a local, real-space invariant for higher-order topological phases. Altogether, the theory we present here provides the groundwork for classifying the topology of metals and other gapless systems of any dimension and in any symmetry class \cite{loringPseuspectra, ryu2010topological} across a broad range of physical platforms.

The remainder of this paper is organized as follows. First, in Sec.\ \ref{sec:review} we provide a brief, physically motivated review of the spectral localizer and Clifford pseudospectrum, and discuss how they can be used to determine a system's topology and the strength of its topological protection. In Sec.\ \ref{sec:chern} we provide a demonstration of how the spectral localizer can be used to identify both the topology and associated boundary-localized states of a metallic Chern lattice. In Sec.\ \ref{sec:hotm} we demonstrate the generality of this method by identifying the topology of a higher-order topological metal, and derive a local invariant for classifying such systems. Finally, in Sec.\ \ref{sec:disc} we offer some concluding remarks.

\section{Review of the spectral localizer \label{sec:review}}

From a broad perspective, the spectral localizer takes a view of a material's topology that is similar to that of topological quantum chemistry \cite{bradlyn_topological_2017,po_symmetry-based_2017,cano_building_2018}: a material is topologically non-trivial if it cannot be continued to an atomic limit without either closing a gap or breaking a symmetry. But, whereas topological quantum chemistry determines whether given material can be continued to an atomic limit by analyzing its band representations, the spectral localizer seeks to make the same determination by instead using the material's real-space description. This real-space picture of topology is predicated on:

\vspace{5pt}
\noindent \textbf{Definition 1:} A material is in an atomic limit if and only if its Hamiltonian, $H$, commutes with all of its position operators, $X_j$, $[H,X_j] = 0 \;\; \forall j$.

\vspace{5pt}
\noindent Using this definition, the question of whether a material is topologically non-trivial becomes synonymous with whether there is an obstruction to continuing a material's Hamiltonian and position operators to be commuting without closing a gap or breaking a symmetry, enabling one to leverage developments from the study of $C^*$-algebras to make this determination \cite{loringPseuspectra,hastings_topological_2011} (in particular, see Fig.\ 1.1 from Ref.\ \cite{hastings_topological_2011}). We note that this definition of the atomic limit is consistent with previous statements about the real-space behavior of this limit \cite{kitaev2009}.

Over the last decade, the spectral localizer has emerged as a versatile tool for identifying whether given set of matrices can be continued to commuting matrices \cite{loringPseuspectra,LoringSchuBa_even, LoringSchuBa_odd}. For a physical material in $d$ dimensions, the spectral localizer is
\begin{multline}
    L_{\boldsymbol{\lambda} = (x_1,\cdots,x_d,E)}(X_1,\cdots,X_{d},H)= \\
    \sum_{j=1}^{d} \kappa (X_{j}-x_{j} I)\otimes\Gamma_{j} + (H-E I)\otimes\Gamma_{d+1}, \label{eq:loc}
\end{multline}
where $I$ is the identity matrix and the matrices $\Gamma_j$ form a non-trivial Clifford representation, $\Gamma_j^\dagger = \Gamma_j$, $\Gamma_j^2 = I$, and $\Gamma_j \Gamma_l = -\Gamma_l \Gamma_j$ for $j \ne l$. Here, $\kappa > 0$ is a scaling coefficient that ensures $X_j$ and $H$ have compatible units, and $\boldsymbol{\lambda} = (x_1,\cdots,x_d,E) \in \mathbb{R}^{d+1}$ is a choice of position and energy where the spectral localizer is evaluated. There are no restrictions on the choices of $\mathbf{x} = (x_1,\cdots,x_d)$ and $E$ in $\boldsymbol{\lambda}$, these quantities can be chosen to be anywhere inside or outside of the material's spatial and spectral extent. Also, note that the underlying theorems that prove the utility of the spectral localizer currently assume that the system's operators $X_1,\cdots,X_{d},H$ are Hermitian, operate on a finite-dimensional Hilbert space, and represent a system with open boundaries.

Intuitively, the spectral localizer can be viewed as a composite of the eigenvalue equations (which have the form $(M-\lambda)\mathbf{v} = 0$) of multiple not-necessarily-commuting operators using a Clifford representation. However, unlike in typical eigenvalue problems where the eigenvalues, $\lambda$, are quantities that are solved for, the spectral localizer takes $\boldsymbol{\lambda}$ as an input and determines whether the system possesses a state with approximate energy $E$ that is approximately at $\mathbf{x}$. If the spectral localizer possesses an eigenvalue that is sufficiently close to zero, 
\begin{equation}
    \min(|\sigma(L_{\boldsymbol{\lambda}}(X_1,\cdots,X_{d},H))|) \leq 
    \sum_{j=1}^d \Vert[H,\kappa X_j]\Vert
\end{equation}
where $\sigma(L_{\boldsymbol{\lambda}})$ denotes the spectrum of $L_{\boldsymbol{\lambda}}$, $\Vert\cdot\Vert$ is the $L_2$ matrix norm, and it is assumed here that $[X_i,X_j] = 0$, then the physical system supports a state in the vicinity of $(\mathbf{x},E)$. If $L_{\boldsymbol{\lambda}}$ does not possess such an eigenvalue, the system exhibits a local gap at $(\mathbf{x},E)$, i.e., a region in position-energy space that cannot support a state (see \cite[\S II]{cerjan_quadratic_2022} for what are currently the best known estimates on how the spectral localizer predicts state localization). Thus, it is convenient to define the ``localizer gap'' as $\min(|\sigma(L_{\boldsymbol{\lambda}})|)$, which, heuristically, can be viewed as a spatially-resolved band gap.

The ability for the spectral localizer to calculate a quantity similar to a band gap without determining the system's band structure plays a crucial role in using $L_{\boldsymbol{\lambda}}$ to categorize a system's topology. Formally, a set of matrices $\{ M_j^{(0)} \}$ can be continued to some other set $\{ M_j^{(1)} \}$ if a continuous path of matrices can be defined between the two sets, $\{ M_j^{(\tau)} \}$ for $0 \le \tau \le 1$. Assessing a material's topology via continuation is typically done for infinite systems, and requires that every set along the path $\{ X_1^{(\tau)},\cdots,X_{d}^{(\tau)},H^{(\tau)} \}$ must both preserve the system's symmetries and maintain the bulk gap (i.e., the system's band gap if it is periodic and infinite). This process also assumes some locality criteria on $H^{(\tau)}$ and $X_j^{(\tau)}$, such that two sites that are sufficiently far apart cannot be coupled \cite{kitaev2009}. Instead, if $H^{(\tau)}$ and $X_j^{(\tau)}$ represent a finite system (as is necessary to use the spectral localizer), the concept of a bulk gap in the continuation process needs to be replaced in someway. One option is to impose periodic boundary conditions and insist that $H^{(\tau)}$ stays gapped, c.f.\ \cite{kitaev2009}. In the case of open boundary conditions, a concept of a local gap is necessary, as the system may possess boundary-localized states (of trivial or topological origin) that would otherwise obscure the spectrum of a system with an insulating interior.

In particular, the criteria of preserving a local gap in $\{ X_1^{(\tau)},\cdots,X_{d}^{(\tau)},H^{(\tau)} \}$ can be guaranteed by monitoring its localizer gap along this path---a system's topology at $\boldsymbol{\lambda}$ cannot change so long as the localizer gap at that $\boldsymbol{\lambda}$ remains open. Specifically, it has been proven that a symmetry preserving perturbation to the Hamiltonian, $\delta H$, is unable to close the localizer gap at $\boldsymbol{\lambda}$ so long as $\Vert \delta H \Vert < \min(|\sigma(L_{\boldsymbol{\lambda}})|)$, \cite[Lemma\ 7.2]{loringPseuspectra}, and a similar statement can be made about perturbations to the position operators, $\delta X_j$. Moreover, this measure of the strength of the topological protection inherently includes the possibility of correlated disorder that is ``designed'' to defeat the system's topology. As such, in most systems, the localizer gap will underestimate the strength of the system's topological robustness for uncorrelated disorder. There has also been recent progress in understanding the effects of perturbations that only approximately respect the system's symmetry class \cite{DollShcuba_Approx_symm_top_ins}.

With these definitions in place, we can present a complete picture of how the spectral localizer determines a material's topology. Overall, there is a constellation of theorems that dictate how the spectral localizer can be used to assess whether a given set of matrices can be continued to commuting matrices for systems of any physical dimension and in any symmetry class \cite{loringPseuspectra,LoringSchuBa_odd,LoringSchuBa_even}. In general, there will be some property of $L_{\boldsymbol{\lambda}}$ that identifies whether such a continuation is possible at a given $\boldsymbol{\lambda}$ while preserving the system's symmetries and without closing the localizer gap at that $\boldsymbol{\lambda}$; this same property also defines a local topological invariant. If there is an obstruction to finding a continuation to commuting operators, the non-trivial topology at that $\boldsymbol{\lambda}$ is protected against perturbations that do not close the localizer gap. Moreover, as the localizer gap is a continuous (but not smooth) function of $\boldsymbol{\lambda}$, neighboring choices of $\boldsymbol{\lambda}$ in position-energy space possess similarly sized localizer gaps (and thus, topology). Finally, bulk-boundary correspondence is built right into this picture, the localizer gap associated with a topologically non-trivial region of position-energy space \textit{must} close around the perimeter of the system where the topological boundary-localized states have strong support, as far away from the finite material the spectral localizer must exhibit trivial topology.

To provide an example for a specific class of topology, consider 2D systems with broken time-reversal symmetry that may possess a non-zero Chern number. For this symmetry class, the system's operators are $H$, $X$, and $Y$, and the system possesses non-trivial topology for some region in position-energy space if $H-EI$, $X-xI$, and $Y-yI$ cannot be continued to commuting while preserving their Hermiticity. In this case, the property of $L_{\boldsymbol{\lambda}}$ that identifies the possible obstruction is its signature, $\textrm{sig}(L_{\boldsymbol{\lambda}})$, which is the number of its positive eigenvalues minus the number of its negative eigenvalues. In particular:
\begin{enumerate}[noitemsep,nolistsep,label=\arabic*)]
    \item If $H$, $X$, and $Y$ commute, then $\textrm{sig}(L_{\boldsymbol{\lambda}}(X,Y,H)=0$ for any choice of $\boldsymbol{\lambda}$ \cite[Lemma 4]{choi_almost_1988}.
    \item The signature of $L_{\boldsymbol{\lambda}}(X,Y,H)$ cannot change through continuation of $H-EI$, $X-xI$, and $Y-yI$ without closing the localizer gap at $\boldsymbol{\lambda}$ \cite{kato_perturbation_2013}.
\end{enumerate}
Thus, even if $H$, $X$, and $Y$ do not commute, if $\textrm{sig}(L_{\boldsymbol{\lambda}}(X,Y,H)=0$ for every choice of $\boldsymbol{\lambda}$, then the system can be continued to an atomic limit without closing the band gap / localizer gap and thus the system (assuming it is sufficiently large) is topologically trivial everywhere \cite{loring_lifting_1997}. Conversely, if there is a choice of $\boldsymbol{\lambda}$ for which $\textrm{sig}(L_{\boldsymbol{\lambda}}(X,Y,H) \ne 0$, then the system possesses non-trivial topology within the localizer gap surrounding this choice of $\boldsymbol{\lambda}$.

In comparison with traditional theories of topology, the local nature of the topology predicted by the spectral localizer can seem unusual, but this is simply the language required to describe widely appreciated properties of topological systems within a real-space picture. Consider a two band topological insulator. Filling only the lower band results in a system that is not Wannierizable (and cannot be continued to the atomic limit), but if both bands are filled the system becomes Wannierizable again. For the spectral localizer, the first case is represented by choosing $\boldsymbol{\lambda}$ for $E$ within the band gap (and $\mathbf{x}$ in the system's bulk) and finding non-trivial local topology, while the second case corresponds to choosing $\boldsymbol{\lambda}$ with an $E$ greater than the maximum energy of the upper band and finding trivial local topology. In the latter case ($E$ outside both bands), the process of continuing the system's operators to commuting will maintain the localizer gap at that $E$, but will close the localizer gap for energies between the bands. Finally, we note that the process of continuing a system's operators to commuting will generally involve changes to \textit{both} $H$ and $X_j$.

\subsection{The Clifford pseudospectrum}

For completeness and to aid a reader in understanding previous works on the spectral localizer, we provide a brief discussion of how the spectral localizer can be used to calculate a system's Clifford pseudospectrum.
Mathematically, the localizer gap defines the system's Clifford $\epsilon$-pseudospectrum,
\begin{multline}
    \Lambda_\epsilon (X_1,\cdots,X_{d},H) = \\
    \{ \boldsymbol{\lambda} \; | \; \min(|\sigma(L_{\boldsymbol{\lambda}}(X_1,\cdots,X_{d},H))|) \le \epsilon \}.
\end{multline}
A system's Clifford spectrum is given by $\Lambda_0 (X_1,\cdots,X_{d},H)$, i.e., the set of $\boldsymbol{\lambda}$ for which the localizer gap vanishes. Thus, a system's Clifford pseudospectrum is a useful tool for finding surfaces in position-energy space with constant localizer gap.

Intuitively, a system's Clifford pseudospectrum (regardless of the system's topology) can be viewed as a method for constructing an approximate joint spectrum of non-commuting operators. In other words, if $\min(|\sigma(L_{\boldsymbol{\lambda}}(X_1,\cdots,X_{d},H))|)$ is small (relative to the norms of the commutators), the system exhibits an approximate eigenstate that almost diagonalizes all of the operators simultaneously with approximate eigenvalues given by $\boldsymbol{\lambda}$ \cite{cerjan_quadratic_2022,loringPseuspectra}.
But, note that even if $\boldsymbol{\lambda}$ is a member of the system's Clifford spectrum (i.e., $\min(|\sigma(L_{\boldsymbol{\lambda}}(X_1,\cdots,X_{d},H))|) = 0$), that does not imply that there is an exact eigenstate that exactly diagonalizes all of the constituent operators. Finally, the Clifford pseudospectra is not the only tool that can be used to understand the approximate joint spectra of non-commuting matrices, which can be tackled using both traditional two-operator pseudospectra \cite{trefethen_hydrodynamic_1993,trefethen_pseudospectra_1997,trefethen_spectra_2005} (in the present context, these could only be used for 1D lattices), or other constructions of multi-operator pseudospectra \cite{cerjan_quadratic_2022} (i.e., different ways of combining eigenvalue equations that need not use a Clifford representation). However, as the Clifford pseudospectra is computed using $L_{\boldsymbol{\lambda}}$, it is the only (currently known) tool for finding approximate joint spectra that is also related to the system's topology.

\subsection{A note on numerical computation}

We conclude this brief review of the spectral localizer with a few comments about the numerical calculation of its properties. Although the spectral localizer is agnostic to the choice of basis used for its constituent operators, if a material's Hamiltonian is written in a tight-binding basis, the position operators, $X_j$, are simply diagonal matrices that index the $j$th coordinate of each of the lattice sites (different orbitals at the same site have the same position). Thus, in the tight-binding basis, $L_{\boldsymbol{\lambda}}(X_1,\cdots,X_{d},H)$, is a (usually large) sparse Hermitian matrix and the localizer gap, $\min(|\sigma(L_{\boldsymbol{\lambda}}(X_1,\cdots,X_{d},H))|)$, can be efficiently calculated using standard sparse eigenvalue solvers as only a single eigenvalue is needed. The properties of the spectral localizer that reveal the system's $K$-theory only need to be calculated a few times, and only once per region in position-energy space where the localizer gap is large, as these properties cannot change without the localizer gap closing.

Moreover, we note that for at least some of the relevant properties of $L_{\boldsymbol{\lambda}}$, there are significant numerical speedups available. For example, to find a matrix's signature, it is \textit{not} necessary to find its full set of eigenvalues. Instead, as $L_{\boldsymbol{\lambda}}(X_1,\cdots,X_{d},H)$ is Hermitian, one can make use of Sylvester's law of inertia \cite{sylvester_xix_1852,higham2014sylvester}, which states that
\begin{equation}
    \textrm{sig}(L_{\boldsymbol{\lambda}}) = \textrm{sig}(D),
\end{equation}
where $L_{\boldsymbol{\lambda}} = P D P^\dagger$ is the LDLT decomposition of the spectral localizer. Thus, as $D$ is diagonal (or block diagonal in some numerical implementations with 1-by-1 and 2-by-2 blocks), the computational cost of finding $\textrm{sig}(L_{\boldsymbol{\lambda}})$ is entirely dictated by the speed of the sparse LDLT decomposition algorithm, which is, in general, more efficient than finding the full spectrum of $L_{\boldsymbol{\lambda}}$.

\section{Topological Chern metal \label{sec:chern}}

\begin{figure}[t!]
    \centering
    \includegraphics[width=1.0\columnwidth]{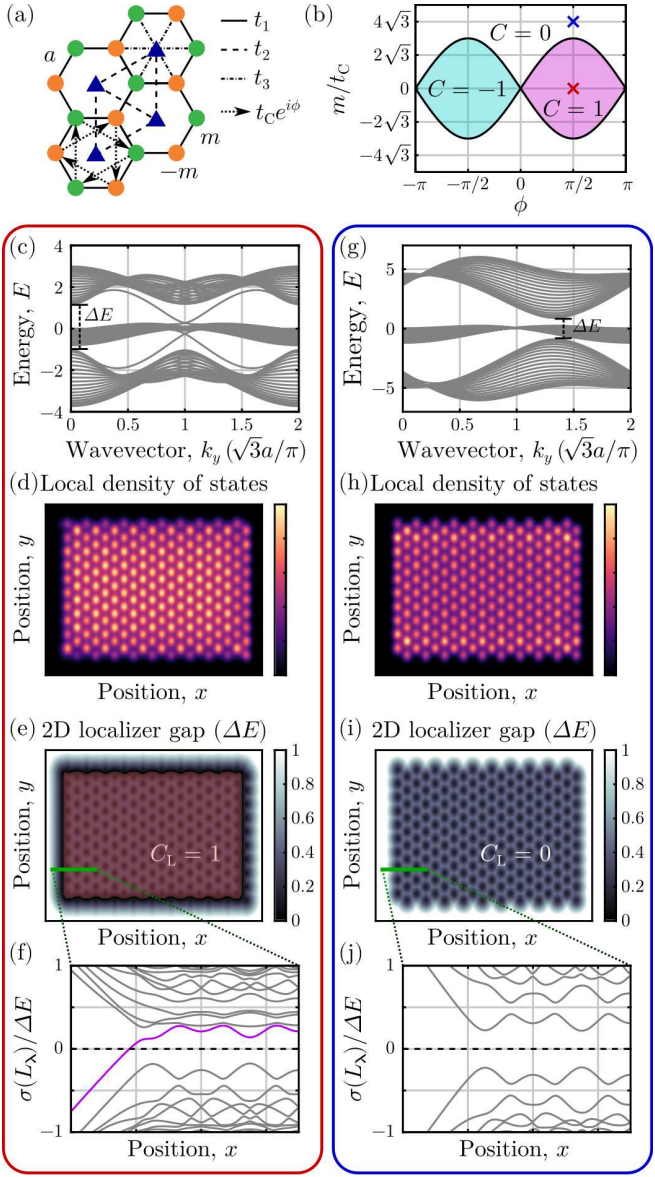}
    \caption{
    (a) Schematic of the tight-binding model for a Haldane lattice, green and orange circles, coupled to a trivial lattice, blue triangles. Some couplings are only shown in a portion of the system for clarity.
    (b) Haldane model phase diagram, with the topological (red) and trivial (blue) systems considered indicated.
    (c-f) Simulations of a metallized Haldane lattice, with $m/t_1 = 0$, $t_C/t_1 = 0.5$, $\phi = \pi/2$, $m_2/t_1 = -0.35$, $t_2/t_1 = 0.2$ and $t_3/t_1 = 0.3$. 
    (c) Ribbon band structure with two zig-zag edges. 
    Chiral edge states can be identified outside of the intervening band. 
    $\Delta E$ is the bulk gap between the top and bottom bands.
    (d) Local density of states at $E = 0$. Each lattice site is represented as a 2D Gaussian with radial width $r_0 = 0.5a$.
    (e) 2D localizer gap, $\textrm{min}(|\sigma(L_{\boldsymbol{\lambda}}(X,Y,H))|)/\Delta E$ at $\boldsymbol{\lambda} = (x,y,E = 0)$ with $\kappa = 1$.
    Overlay shows the local Chern number, $C_{\textrm{L}}(x,y,0) = 1$ (red) or $= 0$ (clear).
    (f) Localizer spectrum along the green line in (e). The eigenvalue which yields a change in topology is highlighted in magenta.
    (g-j) Same as (c-f), except for a trivial metal, with $m/t_C = 4 \sqrt{3}$.
      \label{fig:1}}
      \vspace{-10px} 
\end{figure}

To illuminate how the spectral localizer can be used to classify topological metals, we first consider a 2D Chern insulator with an added intervening band that is degenerate with the Chern insulator's chiral edge states. A minimal tight-binding model for this system can be constructed from a Haldane lattice coupled to a single-band triangular lattice whose vertices are located in the center of each honeycomb \cite{bergman_bulk_2010,junck_transport_2013}, and whose Hamiltonian is
schematically illustrated in Fig.\ \ref{fig:1}a.
The Haldane lattice is parameterized by the nearest neighbor coupling, $t_1$, next-nearest neighbor couplings with amplitude and phase $t_C$ and $\phi$, and the on-site sublattice energy difference $2m$.
The triangular lattice has nearest neighbor coupling $t_2$ and on-site energy $m_2$.
The coupling strength between the two lattices is $t_3$. In the absence of coupling between the two lattices, 
$t_3 = 0$, the Haldane model exhibits topological and trivial insulating phases separated by semi-metal phases and protected by the Chern number, $C$, see Fig.\ \ref{fig:1}b, while the triangular lattice exhibits a single band centered around $E = m_2$. 
When the coupling between the two lattices is turned on, 
$|t_3| > 0$, the intervening band from the triangular lattice prohibits the unique identification of chiral edge states within its extent, as the chiral edge states will generally hybridize with the degenerate states of the interstitial triangular lattice. 

Previous numerical studies have shown that this metallized Haldane model exhibits unusual transport properties that are robust against disorder, even for choices of the Fermi energy that are within the range of the middle band across the entire Brillouin zone (e.g., $E=0$ in Fig.\ \ref{fig:1}c) \cite{bergman_bulk_2010,junck_transport_2013}. However, for choices of the Fermi energy within the middle band, the states responsible for these transport properties cannot be uniquely identified using conventional analysis methods, as any boundary-localized states are degenerate with the bulk states of the triangular lattice. For example, the local density of states (LDOS) at $E=0$ cannot distinguish between topological (Fig.\ \ref{fig:1}d) and trivial (Fig.\ \ref{fig:1}h) metals as the contributions from the degenerate bulk in the system's LDOS outweigh the contributions of any topological boundary-localized phenomena. Moreover, the absence of a bulk band gap at $E=0$ not only poses problems for defining an invariant for this system using band theory, but also inhibits the use of other real-space definitions of topology such as topological markers \cite{bianco_local_chern_2011} as a system's ground-state projection operators are only exponentially localized in gapped systems \cite{kohn_density_1996,resta_kohns_2006,resta_insulating_2011}.

Here, we use the spectral localizer to show that boundary-localized resonances exist in the metallized Haldane model, even for energies within the extent of the middle band, quantify the topological protection of these boundary-localized phenomena, and identify a topological invariant that classifies this behavior. For a 2D system, the spectral localizer can be explicitly written as
\begin{multline}
    L_{\boldsymbol{\lambda} = (x,y,E)}(X,Y,H) = \\
        \left( \begin{array}{cc}
        H - EI & \kappa(X-xI) - i\kappa(Y-yI) \\
        \kappa(X-xI) + i\kappa(Y-yI) & -(H - EI)
        \end{array} \right). \label{eq:loc2d}
\end{multline}
As the spectral localizer directly incorporates information about the system's spatial and spectral properties on equal footing, it is able to identify the approximate presence (or absence) of a state at a given position and energy regardless of other degenerate states elsewhere in the system. Thus, the localizer gap, $\textrm{min}(|\sigma(L_{\boldsymbol{\lambda}}(X,Y,H))|)$, immediately reveals the difference between the topological and trivial phases of the metallized Haldane model. In its topological phase (Fig.\ \ref{fig:1}e), the presence of the chiral edge states causes the localizer gap to close around the entire perimeter of the system regardless of whether the boundary-localized states hybridize with the bulk states of the interstitial triangular lattice. This behavior is qualitatively distinct from the metallized Haldane model's trivial phase (Fig.\ \ref{fig:1}i), where the lack of any boundary-localized states means that the localizer gap remains open around the system's boundary. Moreover, the sizeable localizer gap just inside the boundary closing of the topological system indicates that these boundary-localized states (or resonances) are robust against disorder despite the absence of a bulk band gap. Since the localizer gap is not, in this case, a local manifestation of any sort of a bulk band gap, it is a mathematical mystery why it occurs, related to the discovery in \cite{schulz2021invariants} of how the spectral localizer can exhibit larger gaps than the underlying system in semi-metals.

The topological invariant at any $\boldsymbol{\lambda} = (x,y,E)$ with non-zero localizer gap for 2D systems in symmetry class A is given by 
\begin{equation}
    C_{\textrm{L}}(x,y,E) =\tfrac{1}{2}\textrm{sig}\left( L_{(x,y,E)}(X,Y,H) \right) \in \mathbb{Z}, \label{eq:C}
\end{equation}
Thus, this formulation of the local Chern number is necessarily an integer, even for finite systems. Calculating this invariant for the metallized Haldane model in Figs.\ \ref{fig:1}c-f proves it is topological (i.e., $X-xI$, $Y-yI$, $H-EI$ cannot be continued to be commuting for some $\boldsymbol{\lambda}$), as it exhibits a non-trivial local Chern number in its bulk even for energies residing within the extent of the middle band. Furthermore, this non-trivial bulk topology can be viewed as forcing the localizer gap to close around the Chern metal's entire edge, as the local Chern number must be trivial far away from the system and, thus, the localizer gap along any path connecting the system's interior and exterior must close for one of the localizer's eigenvalues to switch signs, see Figs.\ \ref{fig:1}f,j.

\begin{figure}[t]
    \centering
    \includegraphics[width=1.0\columnwidth]{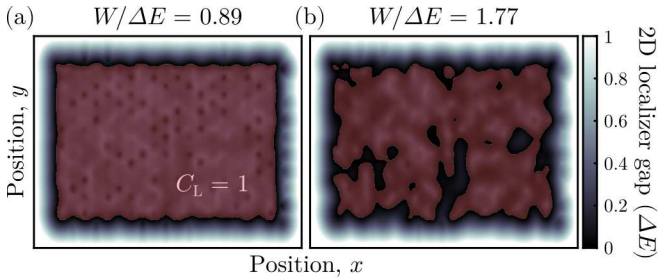}
    \caption{2D localizer gap, $\textrm{min}(|\sigma(L_{\boldsymbol{\lambda}}(X,Y,H))|)/\Delta E$ at $\boldsymbol{\lambda} = (x,y,E = 0)$ with $\kappa = 1$, for a metallized Haldane lattice (same as Fig.\ \ref{fig:1}) with added on-site disorder with strength $W/\Delta E = 0.89$ (a) and $W/\Delta E = 1.77$ (b). The colored overlay shows the local Chern number, $\textrm{sig}(L_{\boldsymbol{\lambda}}(X,Y,H))/2 = 1$ (red) or $= 0$ (clear). Here, $\Delta E$ is the gap between the first and third bands of the system, see Fig.\ \ref{fig:1}c.
      \label{fig:2}}
\end{figure}

As the spectral localizer yields a set of local, real-space definitions for finding boundary-localized states and determining topological invariants, its entire mathematical machinery is immediately applicable in the presence of disorder, without any alteration. Thus, we can show that the topology of the metallized Haldane lattice is robust against perturbations that do not close the gap between the system's first and third bands, $\Delta E$ in Fig.\ \ref{fig:1}c, i.e., those bands which originate from the insulating Haldane lattice. To demonstrate this, we add on-site disorder to the system with strength $W$, such that each vertex has an independent, uniformly distributed random on-site energy in the range $[-W/2,W/2]$. For $W < \Delta E$, the topological character of the system remains unchanged, and the entire bulk still possesses a non-trivial local Chern number (Fig.\ \ref{fig:2}a shows one disorder realization). As the strength of the disorder is further increased, $W > \Delta E$, the system begins to revert to a trivial phase (Fig.\ \ref{fig:2}b). Nevertheless, even at this strength of disorder, regions within the system can still be in a topological phase, and these topological islands can be identified using the local Chern number, Eq.\ (\ref{eq:C}).

\section{Higher-order topological metal \label{sec:hotm}}

To demonstrate the generality of using the spectral localizer to identify topological metals in any symmetry class, we show the existence of robust higher-order topological metallic phases with in-band corner-localized states and find the associated local topological invariant. Moreover, one may argue that in the metallized Haldane model, it is possible to continue the system to a Haldane insulator with an intervening decoupled flat band from the interstitial triangular lattice, enabling the topology of the Haldane insulator to perhaps be inferred from traditional methods. Instead, in this section, we consider a system where the intervening metallic bands are intrinsic to the underlying lattice.

\begin{figure}[t]
    \centering
    \includegraphics[width=1.0\columnwidth]{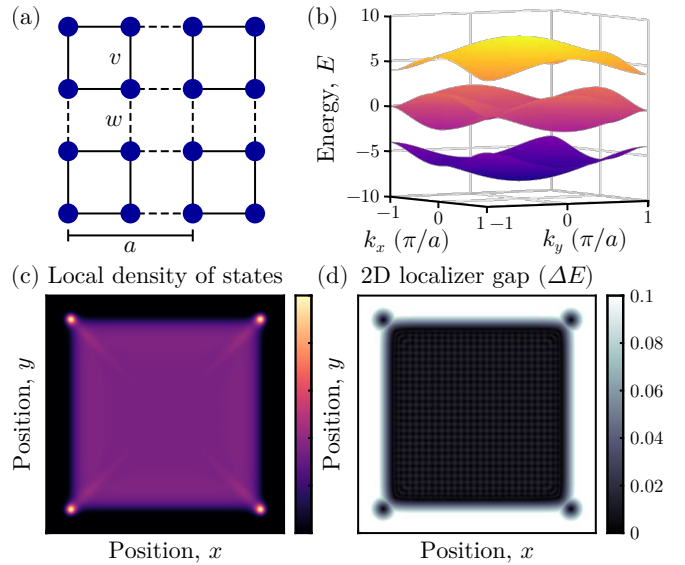}
    \caption{(a) Schematic of the tight-binding model for a higher-order topological metal with intra-unit cell couplings $v$, and inter-unit cell couplings $w$.
    (b) Bulk band structure with $w/v = 3$. $\Delta E/v = 2$ is the bulk gap between the bottom and middle bands.
    (c) Local density of states at $E = 0$. Each lattice site is represented as a 2D Gaussian with radial width $r_0 = 0.5a$.
    (d) 2D localizer gap, $\textrm{min}(|\sigma(L_{\boldsymbol{\lambda}}(X,Y,H))|)/\Delta E$ at $\boldsymbol{\lambda} = (x,y,E = 0)$ with $\kappa = 0.1$.
      \label{fig:3}}
\end{figure}

Here, we consider a 2D chiral and $C_{4v}$ symmetric lattice with four sites per unit cell, whose tight-binding model is schematically shown in Fig.\ \ref{fig:3}a, and in which $v$ and $w$ are the intra- and inter-unit cell couplings, respectively. When a magnetic flux is uniformly threaded through this system, it becomes an insulator at $E = 0$, and when the system is in its topological phase, $w>v$, zero-energy corner-localized states appear \cite{benalcazar2017quad,benalcazar2017quadPRB,song_densuremath-2-dimensional_2017,serra2018observation}. Without this flux, the middle two bulk bands of this system are degenerate and centered at $E = 0$, Fig.\ \ref{fig:3}b. Previous studies of this flux-less system with $w > v$ have shown that so long as $C_{4v}$ (and chiral) symmetry are preserved, corner-localized states exist that are prohibited from hybridizing with the degenerate bulk states \cite{benalcazar2020_HOT_BIC_theory,cerjan2020_HOT_BIC_exp}, and are associated with a non-trivial fractional corner charge invariant \cite{jung_exact_2021}. However, these arguments do not hold in the absence of $C_{4v}$ symmetry, nor do they readily generalize to other gapless systems suspected of exhibiting higher-order topological behaviors. 

We first use the 2D spectral localizer, Eq.\ (\ref{eq:loc2d}), to see that the topological states must remain approximately localized to the corners until any added disorder is strong enough to close the gap between the system's bottom and middle bands (or, equivalently, the gap between its middle and top bands), information which is not available in the system's local density of states, Figs.\ \ref{fig:3}c,d. However, as the metallic system in Fig.\ \ref{fig:3}a is in symmetry class AIII, the 2D spectral localizer is not connected to a non-trivial topological invariant \cite{schnyder2008,kitaev2009,ryu2010topological}. Nevertheless, strong topological invariants in 1D protect 0D edge states, and the 0D corner states of higher-order topological phases are, in the absence of crystalline symmetries, boundary obstructed, rather than bulk obstructed \cite{benalcazar2017quadPRB}. Thus, it seems reasonable to try to treat this system as if it were 1D and borrow the mathematics of the 1D class AIII invariant \cite[\S 4.1]{loringPseuspectra} by projecting the lattice into a lower dimension.

To isolate a corner of the system in the reduced dimension, we use the diagonal position operator $D = (X+Y)/2$.  As a Gedankenexperiment, we are tilting the system and looking in from a corner, and this choice of diagonal position successfully isolates two of the corners. By symmetry, similar behavior will be assured at the remaining two corners. The 1D spectral localizer can be explicitly written as
\begin{multline}
    L_{\boldsymbol{\lambda} = (d,E)}(D,H) = \\
        \left( \begin{array}{cc}
        0 & \kappa(D-dI) - i(H-EI) \\
        \kappa(D-dI) + i(H-EI) & 0
        \end{array} \right), \label{eq:loc1d}
\end{multline}
which allows for the local topological index (assuming non-zero localizer gap) along the diagonal coordinate $d = (x+y)/2$ to be defined as
\begin{align}
    \nu_{\textrm{L}}\left(d,0\right) & = \tfrac{1}{2}\text{sig}\left[ \left(\begin{array}{cc} 0 & I \end{array}\right)
    L_{\left(d,0\right)}(D,H)
    \left(\begin{array}{c} \Pi \\ 0 \end{array}\right)\right] \notag \\
    & = \tfrac{1}{2}\text{sig}\left[ (\kappa(D - dI) + i H) \Pi \right] \in \mathbb{Z}, \label{eq:indAIII}
\end{align}
where $\Pi$ is the system's chiral operator, $\Pi H \Pi = -H$ (and $\Pi D \Pi = D$) \cite{loringPseuspectra,LoringSchuBa_even}. Note, $\nu_{\textrm{L}}$ is \textit{only} well-defined for $E = 0$, which is a mathematical consequence of the fact that the topological states that these chiral-symmetric systems exhibit (if they exist) are guaranteed to be at zero energy. For a true 1D system, $\nu_{\textrm{L}}$ is a local version of the 1D winding number, while for higher-order topological phases in higher dimensional systems we conjecture that it is a local version of the corresponding multipole chiral number \cite{benalcazar_chiral-symmetric_2022} and is related to the infinite-volume invariants in Ref.\ \cite{hayashi2021classification}. In all cases, as $\nu_{\textrm{L}}$ depends on a matrix's signature, it is guaranteed to be an integer, and it is identifying whether $D$ and $H$ can be continued to be commuting while preserving both their Hermiticity and chiral symmetry (and local localizer gap).

\begin{figure}[t]
    \centering
    \includegraphics[width=1.0\columnwidth]{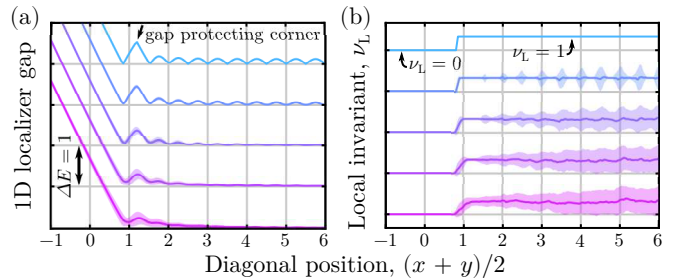}
    \caption{(a) 1D localizer gap, $\textrm{min}(|\sigma(L_{\boldsymbol{\lambda}}(D,H))|)/\Delta E$, for a disordered higher-order topological metal projected along the diagonal position $\boldsymbol{\lambda} = ((x+y)/2,0)$, with $\kappa = 2$.
    Added disorder has strength $W/\Delta E = [0, 0.5, 1, 1.5, 2]$, increasing from cyan to magenta. Solid lines show the average of an ensemble of 100 different disorder realizations, while filled regions show the average $\pm 1$ standard deviation. Data is offset vertically for clarity, each horizontal gray line corresponds to a change in $\Delta E = 1$, and in the bulk of the system $(x+y)/2 > 2$ the ensemble average of the localizer gap is nearly zero for each strength of disorder.
    (b) Similar to (a) for the local topological invariant $\nu_{\textrm{L}}$ given by Eq.\ (\ref{eq:indAIII}).
      \label{fig:4}}
\end{figure}

Calculating Eq.\ (\ref{eq:indAIII}) along the diagonal of the metallic system in Fig.\ \ref{fig:3}a reveals its higher-order topology: the system acquires a non-trivial invariant after the localizer gap first closes at a corner of the system, indicating the presence of a corner-localized state (cyan curves in Fig.\ \ref{fig:4}). Moreover, the nearby localizer gap prevents this state from moving into the system's bulk for disorder strengths $W < 0.5\Delta E$. ($\Delta E$ is the bulk band gap between the top or bottom band and the middle bands.) Thus, even though the corner-localized states will hybridize with the bulk for any strength of disorder, the system's spectral localizer identifies that these states must maintain support on the system's corners until the disorder is strong enough to close the localizer gap. We can explicitly confirm this topological protection by adding disorder to the system which breaks all crystalline symmetries and time-reversal symmetry, where we numerically observe the ensemble averaged localizer gap to remain open and the topological index to remain pinned to $\nu_{\textrm{L}} = 1$ with little variance even for $W = 1.5 \Delta E$, see Fig.\ \ref{fig:4}. This provides numerical evidence for the notion that for uncorrelated disorder the localizer gap is usually an underestimate of the strength of the topological protection in a system. Note that the traces of $\nu_{\textrm{L}}$ in Fig.\ \ref{fig:4}b are ensemble averages, and thus will generally deviate from having an integer value, but each constituent curve in the average is always an integer for any choice of $d$.

\section{Discussion \label{sec:disc}}

In conclusion, we have developed a general theory for assessing a metallic or gapless system's topology using its spectral localizer, even in the presence of disorder. This theory is able to both demonstrate the existence of boundary-localized modes despite a degenerate background continuum, and yields a measure of the strength of these systems' topological protection. To our knowledge, other methods of defining a local or global index all rely on some notion of a gap in the bulk spectrum, perhaps a mobility gap, and are not designed to work in a gapless setting. Indeed, the localizer index was initially designed to work in the presence of a bulk gap, since a bulk gap causes a localizer gap \cite{loringPseuspectra,LoringSchuBa_even, LoringSchuBa_odd}. Nevertheless, we have found that useful localizer gaps can still appear even in the absence of a bulk gap, due to the spatial separation between degenerate states that can be revealed using pseudospectral methods. Although we have only explicitly demonstrated this theory for Chern and higher-order topological metals, this theory should extend without difficulty to all symmetry classes and for systems in any dimension, as the necessary local invariants based on the spectral localizer have already been derived \cite{loringPseuspectra,LoringSchuBa_even,LoringSchuBa_odd}, enabling the discovery of novel topological phases of matter in natural and artificial metals and other gapless materials. 

\begin{acknowledgments}
T.L. acknowledges support from the National Science Foundation, grant DMS-2110398.
A.C. and T.L. acknowledge support from the Center for Integrated Nanotechnologies, an Office of Science User Facility operated for the U.S.\ Department of Energy (DOE) Office of Science, and the Laboratory Directed Research and Development program at Sandia National Laboratories. Sandia National Laboratories is a multimission laboratory managed and operated by National Technology \& Engineering Solutions of Sandia, LLC, a wholly owned subsidiary of Honeywell International, Inc., for the U.S.\ DOE's National Nuclear Security Administration under contract DE-NA-0003525. The views expressed in the article do not necessarily represent the views of the U.S.\ DOE or the United States Government.
\end{acknowledgments}

\bibliography{local_topo_metals_refs}

\end{document}